\documentclass[fleqn,10pt]{wlscirep}
\usepackage[utf8]{inputenc}
\usepackage[T1]{fontenc}
\usepackage{float}
\usepackage{parskip}
\usepackage{notes2bib}

\title{Ultrabroadband THz Conductivity of Gated Graphene In- and Out-of-equilibrium}

\author[1,2,*]{G. Coslovich}
\author[1,3,*]{R.P. Smith}
\author[1,4,5]{S.-F. Shi}
\author[1]{J.H. Buss}
\author[6]{J.T. Robinson}
\author[1,4]{F. Wang} 
\author[1,7,*]{R.A. Kaindl}
\affil[1]{Materials Sciences Division, Lawrence Berkeley National Laboratory, Berkeley, CA 94720, USA}
\affil[2]{Linac Coherent Light Source, SLAC National Accelerator Laboratory, Menlo Park, CA 94025, USA}
\affil[3]{Department of Physics, California State University - East Bay, Hayward, CA 94542, USA}
\affil[4]{Department of Physics, University of California at Berkeley, Berkeley, CA 94720, USA}
\affil[5]{Department of Physics, Carnegie Mellon University,  Pittsburgh, PA 15213, USA}
\affil[6]{U.S. Naval Research Laboratory, Washington D.C., 20375, USA}
\affil[7]{Department of Physics, Arizona State University, Tempe, AZ 85287, USA}
\affil[*]{gcoslovich@slac.stanford.edu, ryan.p.smith@csueastbay.edu, kaindl@asu.edu}

\begin{abstract}
We employ ultrabroadband terahertz (THz) spectroscopy to expose the high-frequency transport properties of Dirac fermions in monolayer graphene. By controlling the carrier concentration via tunable electrical gating, both equilibrium and transient optical conductivities are obtained for a range of Fermi levels.
The frequency-dependent equilibrium response is determined through a combination of time-domain THz and Fourier-transform infrared spectroscopy for energies up to the near-infrared, which also provides a measure of the gate-voltage dependent Fermi level. Transient changes in the real and imaginary parts of the graphene conductivity are electro-optically resolved for frequencies up to 15 THz after near-infrared femtosecond excitation, both at the charge-neutral point and for higher electrostatic-doping levels. Modeling of the THz response provides insight into changes of the carrier spectral weights and scattering rates, and reveals an additional broad-frequency ($\approx$ 8 THz) component to the photo-induced response, which we attribute to the zero-momentum mode of quantum-critical transport observed here in large-area CVD graphene. 

\end{abstract}
\begin{document}
\raggedbottom
\maketitle
\thispagestyle{empty}

\section*{Introduction}

Charge carriers in graphene behave as massless two-dimensional Dirac fermions, yielding unique optical and electronic properties \cite{Novoselov2012}. Importantly, these properties are characterized by very high carrier mobility and strong electromagnetic absorption at THz frequencies. Graphene enables novel applications in diverse fields, including for optoelectronic applications such as photodetectors and communications \cite{Schall2014, Koppens2014}, plasmonics\cite{Grigorenko2012}, light-harvesting\cite{Bonaccorso2015}, ultrafast THz modulation\cite{Tasolamprou2019}, and nonlinear THz effects\cite{Choi2017,Hafez2019}.
Field-induced charge control in graphene has attracted great attention, enabling studies of graphene electrodynamics from low doping into the degenerate regime \cite{Li2008}. While the near-infrared optical transmission of graphene reveals the carriers as massless Dirac fermions through their quantum conductivity\cite{Mak2008}, the optical conductivity in the THz and mid-infrared yields access to frequency-dependent transport determined by intra-band scattering and interband transitions \cite{Li2008,Docherty2012}.
 
Photoexcitation of carriers and the ensuing dynamics ultimately affect graphene's conductivity, which evolves on an ultrafast timescale. Dynamics in graphene has been explored with a range of techniques including mid-infrared and optical pump--probe spectroscopy\cite{Wang2010,Winnerl2011,Malard2013}, optical pump--THz probe spectroscopy\cite{Strait2011,Shi2014,Frenzel2014,Heyman2015,Jensen2014,Tomadin2018}, and time- and angle-resolved photoelectron spectroscopy (trARPES)\cite{Johannsen2013,Gierz2013}. These investigations have provided a window into the physics of non-equilibrium Dirac fermions in graphene, including insights into the mechanisms of energy relaxation due to the interplay of carrier-carrier scattering, impurity scattering, and coupling to phonons. The application of transient THz probes has led to the observation of a sign change in the photoconductivity as a function of the Fermi level. Moreover, the roles of interband and intraband heating in photoexcited graphene have been studied for Fermi levels tuned from near the charge neutral point up to higher energies.\cite{Tomadin2018}

Since the Fermi level plays a key role in transport properties and device physics, electrical gating of graphene has been developed to systematically tune the Fermi energy and thus control the carrier density.\cite{Ren2012}. Such external gating was also applied to time-resolved THz experiments to study the nonequilibrium response of Dirac fermions with different carrier densities and levels of degeneracy \cite{Shi2014,Frenzel2014,Tomadin2018}. However, these first gate-controlled ultrafast THz studies so far measured the response either at the peak of the THz  field or within a limited \textasciitilde 2 THz bandwidth, motivating extension into the multi-THz regime to access higher-frequency contributions to the non-equilibrium transport.

Here, we report ultrabroadband THz studies of gated monolayer graphene, characterizing both its equilibrium conductivity and dynamical response to photoexcitation for a range of Fermi levels. Our phase-sensitive THz measurements enable construction of the real part (direct, in-phase response of currents) as well as the imaginary part (inductive, out-of-phase response) of the graphene conductivity. Equilibrium characterization up to the near-infrared is achieved by combining THz time-domain spectroscopy (THz-TDS) and Fourier-transform infrared (FTIR) spectroscopy. The transient THz conductivity of photoexcited graphene is then probed over an ultrabroadband range up to 15 THz, both at the charge-neutral point and for higher carrier concentrations as controlled by the electrostatic gate. The measurements reveal a distinct broadband ($\approx 8$~THz) component in the equilibrium and photoinduced response, which is modeled through a two-fluid Drude conductivity. This broad component dominates the transient conductivity at the charge-neutral point and contributes fractionally to the photoinduced conductivity changes at higher electrostatic doping. We attribute this response to the zero-momentum mode, previously observed in exfoliated graphene flakes\cite{fengwang2019-2comp}, which arises from quantum critical transport of a Dirac fluid and is reported here for large-area CVD graphene. 

\section*{Results}

\subsection*{Electrical characterization of tunable Fermi level}

In the experiments, we investigate monolayer graphene grown on copper foil via chemical vapor deposition (CVD), which was transferred onto polycrystalline CVD diamond to allow for ultra-broadband THz transmission studies. The graphene sheet covers approximately half of the surface, enabling repeatable characterization of its equilibrium transmission relative to the bare substrate. Raman microscopy shows features confirming single-layer graphene with low defect density (see Methods). The diamond substrate is broadly transparent from the THz to beyond the near-infrared pump energies, unlike other commonly used materials such as sapphire which is limited to \textasciitilde 2 THz. Additionally, unlike small monolayer flakes, the large area of our CVD-grown graphene sheet easily accommodates the diffraction limited focus diameters of long-wavelength THz radiation.

\begin{figure} 
\centering
\includegraphics[width=\linewidth]{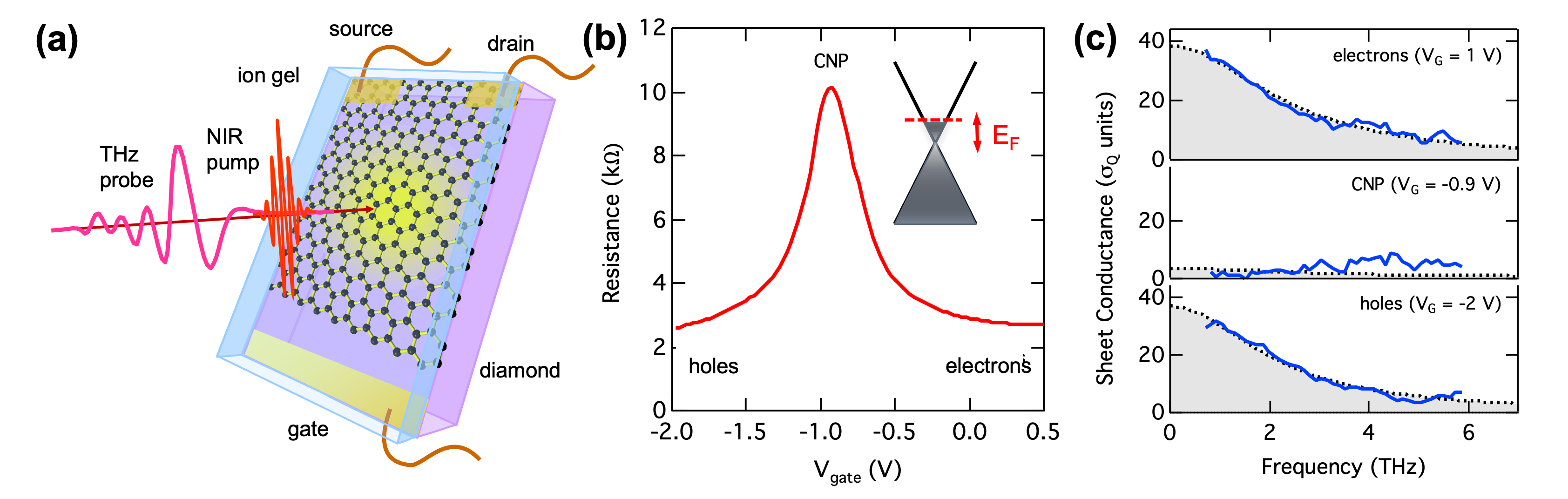}
\caption{Experimental scheme and equilibrium THz response of electrostatically gated graphene. (a) Schematic sample geometry: a thin ion gel layer is deposited on top of graphene and gold contacts for source and drain, while the gate provides a controllable voltage bias and capacitively couples to the graphene monolayer via the ion gel.  Near-infrared pump and THz probe pulses are also illustrated.  (b) Sample resistance vs. gate voltage $V_G$ with electrostatic-doping extremes labeled (CNP: charge-neutral point). Inset: Fermi level $E_F$. (c) Real part of the sheet conductance $\sigma_1(\omega)$ (solid lines) at room temperature for different $V_G$, in units of the quantum conductivity $\sigma_Q$. The response is fitted to a Drude model (dotted) as described in the text, representing the low-frequency intra-band conductivity of Dirac fermions.}
\label{fig:fig1}
\end{figure}
For electrical gating, Au pads were evaporated onto the graphene monolayer and bare substrate, and the structure was subsequently coated with a thin layer of ion gel as described in Ref.~\citen{Shi2014}. The ion gel provides the dielectric layer for the electrostatic gating of graphene. Figure \ref{fig:fig1}(a) illustrates the sample geometry, showing the graphene layer on diamond with Au contacts and the ion gel layer. In Fig. \ref{fig:fig1}(b), we show the measured gate-dependent d.c. resistance of the electrically charged CVD graphene. Measured at a small bias voltage of $\approx$50 mV, the resistance varies continuously and reaches a maximum at the charge-neutral point (CNP) where electron and hole populations are balanced.

\subsection*{Equilibrium THz characterization} \label{sec:equil} 

We first determine the low-frequency equilibrium conductivity of our gated graphene sample for various Fermi levels using broadband time-domain THz spectroscopy. This approach employs optical rectification and electro-optic detection with thin GaP crystals, covering the 1--6 THz (33--200 $\mathrm{cm}^{-1}$) range. Figure \ref{fig:fig1}(c) shows the real part of the optical sheet conductivity of the monolayer for different gate voltages, as obtained from the complex-valued THz field transmission at room temperature. At the CNP, the THz conductivity is strongly suppressed. In contrast, at high gate voltages the THz spectra reveal a strong Drude peak, indicative of the intra-band conductivity of a dense Dirac plasma. To inject a high concentration of electrons into our graphene, we adjust the gate voltage to +1.0 V, thereby tuning the Fermi level up to +298 meV (corresponding to a carrier density of $\approx 7 \times 10^{12} cm^{-2}$). We can inject similar concentrations of hole carriers by adjusting the gate voltage to -2.0 V, resulting in a Fermi level of -255 meV.

We fit the equilibrium graphene conductivity with a model that includes both intra- and inter-band contributions, $\sigma(\omega) = \sigma_{\mathrm{intra}}(\omega) + \sigma_{\mathrm{inter}}(\omega)$. The intraband conductivity describes a single-component Drude transport:
\begin{equation}
\sigma_{\mathrm{intra}}(\omega) = \frac{\omega_p^2}{4 \pi } \frac{1}{\omega^{2} \tau + 1 / \tau}\left ( 1+ i \omega \tau \right ) ,
\label{eq:Drude_singleIntra}
\end{equation}
where the plasma frequency $\omega_p$ is given by $\omega_p^2= 4 \sigma_Q \frac{8 k_{B} T}{\hslash} \mathrm{ln}\left ( e^{-E_{F}^{e}/2 k_{B} T} + e^{E_{F}^{e}/2 k_{B} T}\right ) $ and $\sigma_Q~=~\frac{e^2}{4 \hslash}$ is the quantum conductivity.\cite{Choi2009} The interband conductivity includes a real part,
\begin{equation}
\Re \sigma_{\mathrm{inter}}(\omega)= \frac{\sigma_Q}{2}\left [ \mathrm{tanh}\left ( \frac{\hslash \omega + 2 E_{F}^{e}}{4 k_{B} T} \right ) + \mathrm{tanh}\left ( \frac{\hslash \omega - 2 E_{F}^{e}}{4 k_{B} T} \right )\right ],
\label{eq:Drude_singleInter}
\end{equation}
while for the imaginary part of the interband term we use the low-temperature formula from Falkovsky et al.\cite{PhysRevB.76.153410} as an approximation.
To simulate the sample transmission, we utilize a transfer matrix calculation\cite{Kaindl2009} that incorporates the substrate, graphene layer, and ion gel layer in our system (see Methods).

As evident from Fig.~\ref{fig:fig1}(c), in the high-doping regime the real part of the single-component Drude response fits the data well over the 1--6 THz frequency range. We extract similar scattering rates in the electron-doped and hole-doped cases of $1/\tau_e \approx 2.3$ THz and $1/\tau_h \approx 2.1$ THz, respectively. The Fermi level values obtained from this model are, correspondingly, $E_{F}^{e}\approx 300$ meV and $E_{F}^{h}\approx 255$ meV. In contrast, for the CNP equilibrium data, only a very small intraband conductivity is observed, as expected when $E_{F}$ approaches zero. No reliable scattering rate could be extracted at this doping due to the small signal levels. 
While small deviations from a conventional Drude response may be observed on the high frequency side of the equilibrium THz spectra \cite{Whelan_2021,PhysRevB.64.155106}, we retained a Drude lineshape for the purpose of our analysis.
\bibnote{we attempted to fit our CNP data using a Drude-Smith model (not shown), which includes back-scattering effects due to inhomogeneities and confinement. The Drude-Smith model produced slightly better agreement with the data for the CNP case, yet still without a reliable extraction of parameters. Thus, for simplicity, we adopted a single-Drude model as a starting point for our analysis.}

\subsection*{FTIR broadband spectroscopy results}

To characterize the gate-tunable equilibrium conductivity of the graphene monolayer at even higher frequencies -- accessing the range of interband transitions -- we employed FTIR spectroscopy.
For this, room temperature transmission spectra were recorded in the mid- and far-infrared covering the 20--180 THz spectral range (corresponding to $\approx$~670-6000 cm$^{-1}$).

To sensitively determine the broadband transmission through the thin graphene monolayer, we compute the ratio of the FTIR transmission spectrum -- obtained at the same sample location -- between the electrically-doped and CNP cases. This method, introduced by Li et al.\cite{Li2008}, is critical to characterize the small transmission differences in graphene which are on the order of \textasciitilde 1\%, since moving to a different location on the sample typically changes the magnitude of transmission by a comparable amount. This approach allowed us to measure infrared spectra with sensitivity to the small transmission changes arising from the electrically-tuned doping levels. 

\begin{figure}[H]
\centering
\includegraphics[width=\linewidth]{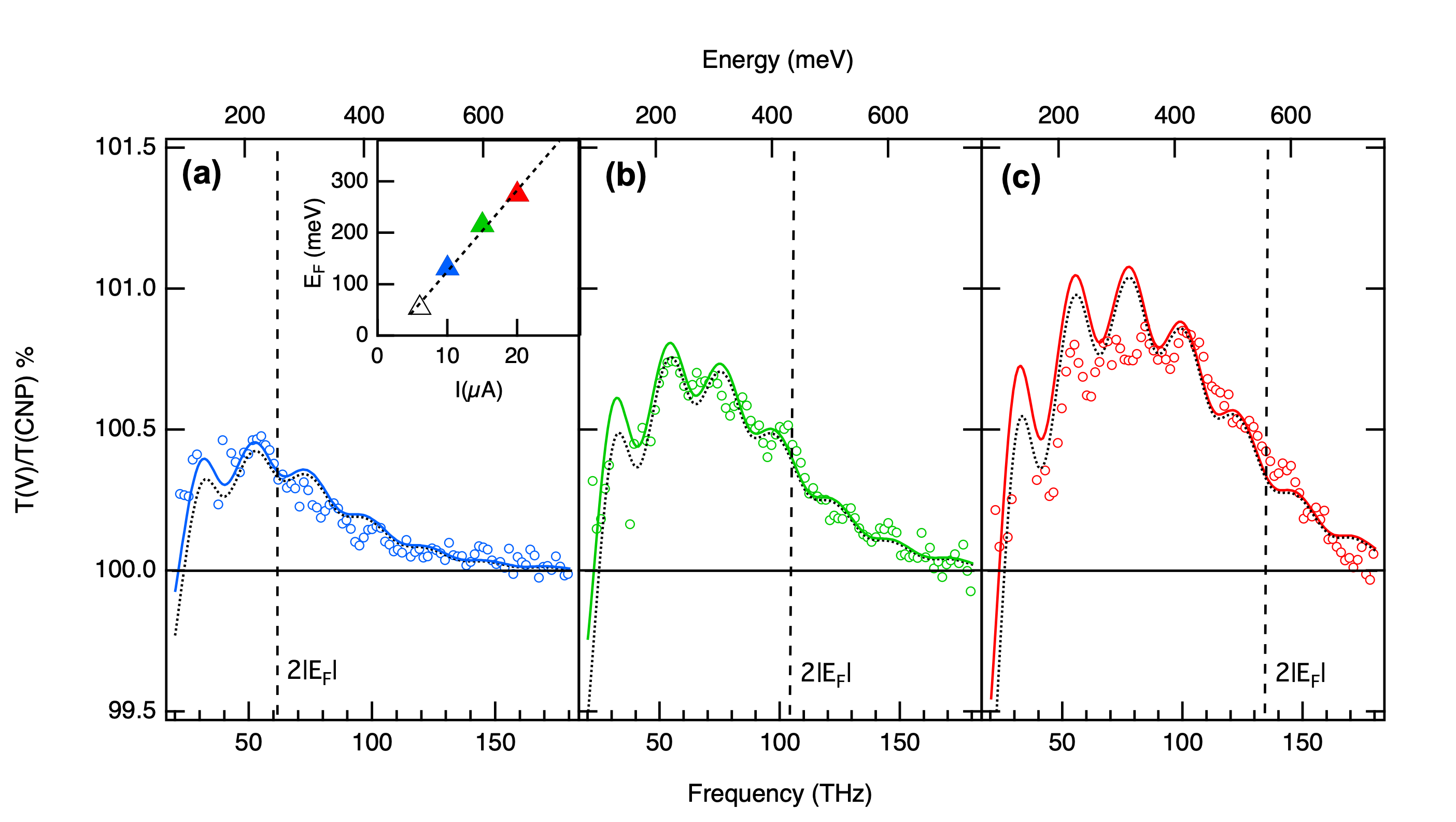}
\caption{
FTIR transmission ratio of the graphene monolayer gated on the electron-doped side to that at the charge-neutral point (CNP). Electron doping corresponds to (a) $E_F =$ 130 meV, (b)  $E_F =$ 215 meV, and (c)  $E_F =$ 275 meV, while the CNP doping level is at $E_F =$ 50 meV. Vertical lines in each panel indicate the inflection points at $2E_F$. Inset: linear relationship between the source-drain current and the Fermi level. Data is shown as circles, alongside single-Drude (solid line) and two-component Drude model (dotted line) fits. 
The vertical dashed lines in each panel indicate $2 E_F$ for each of the three doping levels.
}
\label{fig:equil_fig}
\end{figure}

Figures~\ref{fig:equil_fig}(a--c) show the resulting transmission-ratio spectra for three representative carrier concentrations, with the gate tuned to the electron-doped case ($E_F=$ 130, 215, and 275 meV) at room temperature. Similar FTIR transmission curves were observed for equivalent hole-doped concentrations (not shown).
We fit the FTIR data for each doping level with the above model [Eqs. ~(\ref{eq:Drude_singleIntra}) and (\ref{eq:Drude_singleInter})] that takes into account a single-component Drude transport and interband transitions, with the transmission ratios calculated using the transfer matrix calculation described in the Methods section. 
Fits are shown as solid lines in Figs.~\ref{fig:equil_fig}(a)-(c). The Fermi level values were obtained from this fit and confirmed by visual inspection of the inflection points in the infrared spectra, indicating the drop in interband conductivity at $2E_F$.

This simple model fits the data reasonably well, with however some discrepancies on the low-frequency side at the higher doping concentrations.
Deviations from a single-component Drude model of itinerant carriers are known for materials with many-body effects\cite{BasovRevModPhys.83.471}, where an extended-Drude model framework has been successfully applied\cite{Basov2014}.
In our data, the deviations suggest the presence of a broader spectral component, as discussed further below.

\subsection*{Transient conductivity in gated graphene}

Having characterized the equilibrium response, we investigate the dynamical behavior of photoexcited carriers in gated graphene.
The monolayer is excited with near-infrared 800 nm pulses, whose photon energy exceeds $2 E_F$ and thus promotes electrons into the graphene conduction band and creates a nonequilibrium state. To probe the THz-frequency transport and ultrafast dynamics of non-equilibrium photoexcited carriers in graphene, we performed optical-pump THz-probe (OPTP) spectroscopy at different doping levels, taking advantage of the ultra-broadband THz frequency range of our setup. In the experiments, the photoexcitation fluence was varied between 5 and 50 $\mu \mathrm{J}/\mathrm{cm}^2$, yielding access to different excitation regimes relative to the equilibrium carriers. In all experiments reported here, the near-infrared pump and THz probe polarizations are linear and aligned parallel. The transient change of the transmitted THz electric field $E(t)$ was obtained for frequencies up to 15 THz by combining data from two sets of THz detector/emitter pairs, comprised of either GaP (1--6 THz) or GaSe (9--15 THz) crystals.

\subsubsection*{Temporal dynamics}

\begin{figure}[H]
\centering
\includegraphics[width=\linewidth]{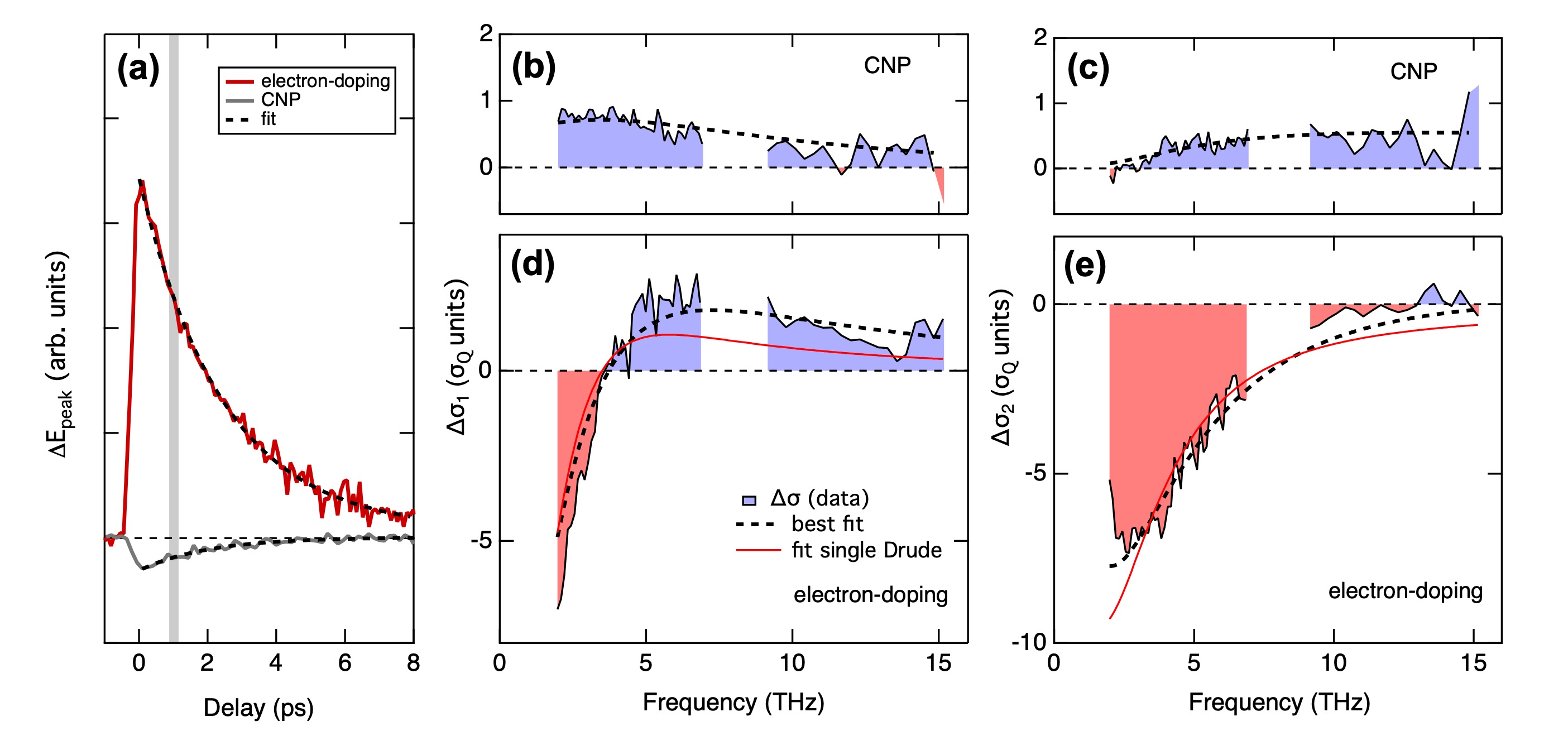}
\caption{Ultrabroadband THz dynamics of graphene, for the electron-doped and charge-neutral point cases. (a) THz field change $\Delta E(t)$ at the THz pulse peak (GaP detection) for 800 nm photoexcitation with 50 $\mu \mathrm{J}/\mathrm{cm}^2$ fluence. The dotted lines are exponential fits of the relaxation dynamics.
Photoinduced change of the (b,d) real part and (c,e) imaginary part of the THz conductivity $\Delta \sigma(\omega)$ at 1 ps pump-probe delay. The red lines are the single-Drude fits to the data, and the black dashed lines are the best fits with a two-component Drude model.
}
\label{fig:time-res-overview}
\end{figure}

Fig. \ref{fig:time-res-overview}(a) shows the dynamics of the pump-induced field change $\Delta E$ at the THz pulse peak, measured with GaP detection up to 6~THz. Around the charge-neutral point (CNP) $\Delta E$ exhibits a negative sign, corresponding to increased conductivity. This results from the added intraband spectral weight arising from the photo-excited hot carriers. However, we observe a positive photoinduced $\Delta E$ when the sample is gated to the high-density  electron or hole side. This case corresponds to a conductivity decrease, compatible with a broadened Drude response with higher scattering rate. These qualitative observations are in agreement with previous studies in the few-THz range\cite{Shi2014,Frenzel2014,Heyman2015,Jensen2014,Tomadin2018}. The transient THz signals relax back to equilibrium with a \textasciitilde 2~ps single-exponential decay, which we find not to depend significantly on the field-induced Fermi levels. 
We subsequently investigate the transient spectra of the ultrabroadband THz response to gain insight into the spectral origin of this ultrafast conductivity dynamics in photoexcited graphene.

\subsubsection*{Spectral dependence of the complex conductivity}

Spectra of the pump-induced sheet conductivity $\Delta \sigma_1(\omega)$ are shown in Figs. \ref{fig:time-res-overview}(b)--(e) at 1 ps pump-probe delay, providing an ultrafast snapshot of the ultrabroadband THz transport properties of non-equilibrium Dirac fermions. The time delay is referenced to the peak of $\Delta E(t)$ shown in Fig. \ref{fig:time-res-overview}(a). The transient data, spanning the 2-15 THz range, reveal novel features of graphene transport hidden in the low-frequency response. As shown in Figs.~\ref{fig:time-res-overview}(b) and (c), optical perturbation of the graphene monolayer around the CNP results in an induced broadband Drude conductivity extending up to \textasciitilde 15 THz. From the single-Drude fit, the $\Delta \sigma_1(\omega)$ response shows a width of $\approx$8~THz with a corresponding peak in $\Delta \sigma_2(\omega)$ in the same frequency range. This observation indicates a clear dichotomy, with the carriers  optically induced at the CNP experiencing scattering rates far exceeding those of the electrostatically-doped carriers in equilibrium ($\approx$2-3 THz spectral width).

In contrast with the CNP case, as shown in Figs.~\ref{fig:time-res-overview}(d) and (e) in the high-doping (degenerate) regime the photo-induced conductivity changes $\Delta \sigma_1(\omega)$ are strongly frequency-dependent and change sign as the frequency exceeds about 3 THz. This response can be attributed to both a marked spectral weight loss of the equilibrium intraband term and a strongly-enhanced scattering rate of the non-equilibrium carrier distribution. 

The qualitative trend of these curves can first analyzed by extension of the standard single-component Drude model (Eqs. \ref{eq:Drude_singleIntra} and \ref{eq:Drude_singleInter}) to the non-equilibrium case. The perturbation of the conductivity following photoexcitation is $\Delta \sigma(\omega) = \sigma_{\mathrm{exc}}(\omega) - \sigma_{\mathrm{eq}}(\omega)$, and we vary the free parameters starting from the equilibrium model to attempt a best fit of the transient data. In the CNP case, the transient conductivity is indeed well reproduced by the appearance of a new $\approx$8 THz-wide Drude peak. 

In the high-doping case we consider first a single-Drude analysis, with a change of scattering rate ($1 / \tau$) and of $E_{F}$ from their equilibrium values. The best fits in this case are shown as red lines in Figs.~\ref{fig:time-res-overview}(d) and (e). From the fit we obtain $1/\tau_{\mathrm{exc}} \approx~3.5$ THz and ${E_{F}^{e}}_{\mathrm{exc}}\approx~280$ meV with an effective electronic temperature around 700 K. However, as evident in Figs.~\ref{fig:time-res-overview}(d) and (e) this single-component Drude fit falls significantly short of reproducing the photo-induced THz conductivity over the extended frequency range.

\section*{Discussion}

Considering the failure of the single-Drude model to accurately match our transient conductivity measurements, we are in need of a different model that will faithfully reproduce the data. In particular, the experimental spectra evidence the appearance of a higher-frequency spectral weight which can be attributed to an additional broader conductivity component contributing to the photo-excited spectrum. To gain further insight into the physics underlying the broadband transient conductivity response we thus explore the application of a multi-component response. Previous studies have highlighted instances where a conventional, single-component Drude model falls short in describing the transport properties of quantum materials\cite{Gierz2013,Nakajima2014,Mihnev2016}.

In particular, recent experiments using narrowband, on-chip THz spectroscopy of high quality, hBN-encapsulated monolayer graphene flakes have evidenced the existence of two distinct transport modes of its Dirac carriers \cite{fengwang2019-2comp}. In this picture, the transport near the CNP corresponds to a 'zero-momentum mode' of a Dirac fluid, where the counterpropagating motion of the balanced electron and hole populations results in mutual charge carrier collisions at a quantum critical rate of $\approx k_B T/\hslash$.\cite{PhysRevB.78.085416} With increased doping, a 'finite-momentum mode' appears that corresponds to the transport of a co-propagating, charge-imbalanced carrier fluid, with a completely different rate dominated by impurity and phonon scattering. The relative spectral weight of the two components depends on the level of gate-controlled electrostatic charge doping \cite{fengwang2019-2comp}.

To investigate whether this picture can explain our results in monolayer CVD graphene, we analyze the ultrabroadband response with a two-fluid conductivity. In this model, the interband term $\sigma_{\mathrm{inter}}(\omega)$ remains unchanged and follows Eq. (\ref{eq:Drude_singleInter}), while the intraband term is now given by two transport modes as $\sigma_{\mathrm{intra}}(\omega) = f\sigma_{\mathrm{intra},1}(\omega) + (1-f)\sigma_{\mathrm{intra},2}(\omega)$. Each mode corresponds to a Drude conductivity
\begin{equation}
\sigma_{\mathrm{intra},i}(\omega) = \frac{\omega_p^{*2}}{4 \pi } \frac{1}{\omega^{2} \tau_i + 1 / \tau_i}\left ( 1+ i \omega \tau_i \right ) ,
\label{eq:Drude_ext}
\end{equation}
with $\tau_i=1/\gamma_i$ and $\omega_p$ as described in Ref.~\citen{Basov2014}. This model thus describes the Drude-like transport of a mode with spectral weight $f$ and scattering rate $\gamma_1$, and one with $(1-f)$ spectral weight and scattering time $\gamma_2$. 


\subsubsection*{Fitting equilibrium data using a two-component Drude model}

We first apply the two-fluid model to the measured FTIR spectral transmission ratios between gate-voltage doped and CNP states. While the single-Drude model captures most of the features, we examine whether the deviations can be addressed. As shown in Fig.~\ref{fig:equil_fig} the two-component Drude model (dotted lines) follows the data more closely than the single Drude model (solid lines), especially in the lower spectral range of Fig.~\ref{fig:equil_fig} (c). This fit was obtained by iterative optimization of the two-component model (Eq.~\ref{eq:Drude_ext}) on the FTIR transmission ratios and THz conductivity.  

In the model, the first component was initially set to the parameters of the single-Drude fit, while the second component was set to $\gamma_2 \approx $ 8 THz to match the scattering rate of the transient carriers photoexcited into the CNP state (Fig.~\ref{fig:time-res-overview}). To maintain a good fit of the THz equilibrium data, the scattering rate of first, narrowband component was slightly adjusted to $\gamma_1 \approx$ 2.2 THz. We find that for the range of doping concentrations studied, a relative spectral weight of $f=$~55\% of the carriers being low scattering rate population and 45\% of the carriers as high scattering rate population enabled a closer description of the FTIR equilibrium data, while remaining compatible with the equilibrium THz-TDS data in Fig \ref{fig:fig1}. 

The high scattering component we observe matches (in terms of scattering rate and spectral weight) the properties of the zero-momentum Dirac fluid reported in Ref.~\citen{fengwang2019-2comp}, suggesting an origin of this broad component rooted in the quantum critical transport of graphene. Its impact on the FTIR transmission ratios is larger than on the THz-TDS data due to the sensitivity of the high-frequency response to small changes of the broadband component. 

\subsubsection*{Fitting transient data using a two-component Drude model}

Having seen evidence for a two-fluid model being relevant to the equilibrium response, we next apply this model to the transient THz conductivity. As shown in Figs. \ref{fig:time-res-overview}(d) and (e), the two-component fits (dashed lines) simultaneously match both real [Fig. \ref{fig:time-res-overview}(d)] and imaginary [Fig. \ref{fig:time-res-overview}(e)] parts of the ultrabroadband transient conductivity spectra. This contrasts with the fit of a single-component Drude transport with transient changes of its scattering rate and spectral weight (red lines), which deviates significantly. 

The close agreement of our data with the two-fluid model thus suggests a physical picture in which two carrier modes with distinctly different scattering rates contribute to the THz transport. The additional transport resulting from photoexcitation exhibits high scattering rates, leading to a broad feature in the spectrum which remains for several picoseconds after excitation. Simultaneously, the spectral weight of the existing narrowband mode is reduced, resulting in a spectrally sharp feature in the difference spectra $\Delta\sigma_1$ at lower frequencies. 

Since we observed the broad component as the sole photo-excited component at the CNP, we argue that the high-scattering response ($\gamma_2$) is representative of the zero-momentum mode expected to dominate CNP transport in graphene\cite{fengwang2019-2comp,PhysRevB.78.085432}. Within our two-fluid framework, the response observed for the electrostatically-doped graphene then results from the carrier spectral weight shifting from the narrowband finite-momentum to the broadband, zero-momentum mode. Indeed, the fits reveal optical pumping to modify the spectral weight balance, with the narrowband mode ($f$) changing from 55\% to 35\% and correspondingly the broadband (zero-momentum) mode  shifting from 45\%  to 65\%. These shifts explain the observed changes of both  real and imaginary parts of the transient THz conductivity. 


It should be noted that our time-resolved probe is particularly sensitive to small changes in the THz response, on the order of the quantum conductivity $\sigma_Q$ or less, unlike the equilibrium THz-TDS data where noise and calibration issues can affect the sensitivity on the $\sigma_Q$ scale. The deviations due to the two-fluid model are particularly visible at higher frequencies (> 10 THz) motivating the use of ultrabroadband THz probes. We expect that additional temperature and fluence dependent studies will allow to further clarify the behavior of the Dirac quantum fluid and the limits of the zero-momentum mode picture in CVD graphene.

\section*{Conclusion}

In summary, our work represents the first time-resolved ultrabroadband THz study  of gated graphene following optical excitation. The equilibrium and transient THz conductivity over an extended frequency range yield direct access to dynamical changes of the scattering rates and spectral weight, delivering novel insight into non-equilibrium transport of Dirac fermions. A two-fluid Drude model closely reproduces both the equilibrium and transient THz conductivity. We identify a photoinduced high-scattering-rate component as the zero-momentum mode of the Dirac fluid in graphene, where electrons and holes propagate in opposite directions and scatter at quantum-critical rates. This mode was previously reported in THz studies of exfoliated monolayer graphene \cite{PhysRevB.78.085416,fengwang2019-2comp}, while our measurements expand its observation and applicability to large-area, monolayer CVD graphene.
The THz probe resolves transient spectral weight shifts between low and high scattering-rate modes triggered by photoexcitation. These results motivate additional time-resolved ultrabroadband THz and ARPES experiments to sensitively detect the zero-momentum mode and elucidate its correlation with transient distribution functions in graphene. Moreover, we expect the combination of gate-tuning and ultrabroadband THz probes to enable critical insight for a variety of novel systems such as 'magic-angle' twisted bilayer graphene\cite{Cao2018} and few-layer transition metal dichalcogenides\cite{Lu2018}.

\section*{Methods}

\subsection*{Sample fabrication}

We use graphene grown by CVD onto copper foil and transferred onto a 0.5-mm thick polycrystalline CVD diamond substrate (Element Six Diafilm OP, 8-mm diameter, polished both sides, Ra < 30 nm).  Before the transfer, a 90-nm SiO$_2$ layer was evaporated onto one side of the diamond substrate. This SiO$_2$ layer\cite{Blake2007} improves the visibility of monolayer graphene under an optical microscope, which assists experimentally in finding the appropriate spot on the sample to optically probe. The graphene sample covers approximately half of the substrate area, to allow for repeatable characterization of the equilibrium graphene layer conductivity. The utilization of a diamond substrate is critical to the broadband THz experiments as it is one of few materials that is transparent to both the optical pump and high-frequency THz radiation\cite{Kubarev2009} above \textasciitilde 6 THz.

\subsection*{Ion-gel gating of graphene}

To control the Fermi levels of graphene on diamond systematically, we then apply Au contact pads followed by a coating by an ion gel. The Au pads were thermally evaporated directly onto the graphene sample, using a 5-nm Cr sticking layer. 
We drop an ion gel\cite{Chen2011,Shi2014} (triblock PS–PEO–PS copolymer) onto our graphene and allow it to dry for several days. 
This fabrication results in a field-effect transistor, where the Fermi-level of the graphene can be systematically tuned by application of a voltage at the gate terminal, as shown in Fig. \ref{fig:fig1}a, similar to Tomadin et al.\cite{Tomadin2018}.
Using this configuration, the doping concentration can be controlled through a range of electron-doped and hole-doped concentrations as well as close to the charge-neutral point (CNP). The finite sheet resistance at the CNP voltage, corresponding to a Fermi level of 40 meV, is attributed to unavoidable charge impurities in our sample. 
The ion gel, while necessary for conveniently controlling the doping concentrations in our sample, is also an optical layer (index of refraction $n= 1.35$) which the THz field must pass through. 

To measure the resistivity curve shown in Fig. \ref{fig:fig1}b, we applied a 50 mV bias between the source and drain electrodes while scanning the gate voltage.

In our OPTP measurements, we verified that there is no transient signal in the ion gel on the same substrate in the absence of graphene.

\subsection*{Verifying quality of single-layer graphene sample}

To confirm the quality of our sample, we performed Raman characterization of the graphene sample with a Bruker SENTERRA RFS100/S Raman spectrometer using a 785 nm laser under ambient conditions.
Raman spectroscopy confirms the high quality of our single-layer graphene sample covering half of the substrate. The ratio of the 2D/G peaks near 2 consistently across the graphene verifies the consistent single layer, and the very small D-peak intensity indicates a low defect density\cite{Basov2014}.

\subsection*{Electrical transport measurements}

We characterize the electrical transport characteristics of our sample in situ by applying a gate voltage in both forward and reverse biased polarities over a range of +/- 5 Volts. By recording the resistance across the source and drain of our sample using a Keithley 196 digital multimeter, we could conveniently tune the Fermi level to the desired level for each experiment.

\subsection*{Optical-pump THz-probe (OPTP) setup}

The time-domain and ultrafast THz experiments are driven by a 250-kHz regeneratively amplified Ti:sapphire laser system. A portion of the 800 nm, 50-fs laser pulses is used to generate THz pulses by difference-frequency generation in either ZnTe (0.5--3 THz), GaP (1--6 THz), or GaSe (9--15 THz) crystals.  The generated THz radiation is collimated and focused onto the sample using parabolic mirrors.  After passing through a 200 $\mu$m pinhole followed by the sample, the transmitted THz light is recollimated and focused onto an electro-optic detection crystal (which is matched to the emitter crystal).  

For the time-resolved experiments a portion of the 800 nm laser output is used to pump the graphene sample from the opposite direction, at a variable time delay before the THz pulse arrives at the sample. The pump focus diameter is set to \textasciitilde 500-$\mu$m full-width at half-maximum (FWHM). The photoexcitation densities are controlled by filtering the pump beam fluence  with a variable neutral density filter. The probe delay is typically set to 1 ps or longer, exceeding the thermalization time of the electronic distribution, which ensures that we probe the system having reached a quasi-equilibrium state. Equilibrium and transient data are both measured using the same lock-in amplifier for identical geometries.  For electro-optic sampling of the full THz fields $E(t)$ transmitted through the sample, the lock-in amplifier is referenced to a mechanical chopper in the generation arm. For measurements of the pump-induced field change $\Delta E(t)$, the pump beam is chopped with the  generation chopper turned off. All paths of the THz beam are under continuous nitrogen purge starting at least 30 minutes before the measurements, to suppress absorption from water vapor and other gases in the THz spectral range.

\subsection*{Extracting $\sigma(\omega)$ and $\Delta \sigma(\omega)$ from the measured THz fields}

After recording and Fourier-transforming the TDS THz signals for $E(t)_\mathrm{film+subs}$, the THz transmission through a thin film on an insulating substrate, and $E(t)_\mathrm{subs}$, and the THz transmission through the substrate only, we can relate the ratios of the transmission spectra $T=E(\omega)_\mathrm{film+subs}/E(\omega)_\mathrm{subs}$ to the conductivity through $T = \frac{1+N}{1+N+Z_0 \sigma d}$, where $N=2.377$ is the index of refraction of the diamond substrate, $Z_0 =377~\Omega$ is the free-space impedance, and $d$ is the thickness of the thin film (graphene layer)\cite{Kaindl_2012}. We then verified that the extracted $\sigma(\omega)$ using a numerical recursive transfer matrix technique that included the optical layers of our sample \cite{BornWolf:1999:Book}.

To characterize the pump-induced transient THz conductivity, we first measure with a Lock-in amplifier the electro-optic signal from the THz field $E(t)$ transmitted through the unexcited sample, with the near-infrared THz generation beam modulated with an optical chopper. The pump-induced changes of the transmitted THz field $\Delta E(t)$ are then recorded while chopping the pump beam. The two measurements are carried out separately but in direct succession. From this, we obtain the transient conductivity as $\Delta \sigma(\omega) = - \frac{1+N+Z_0 \sigma_{\mathrm{eq}} \sigma_{\mathrm{Q}}}{\sigma_{\mathrm{Q}}} \frac{\Delta E(t)}{E(t)}$. As evident in this equation, the transient conductivity change is proportional to the electric field change but with opposite sign.

For both equilibrium and transient data, we included a transfer matrix calculation reproducing the etalon effect of the ion gel, which we measured to have an index of refraction $n= 1.35$ and average thickness 6.4 $\mu$m in modeling the transmission data. While fits including and not including the effect of the ion gel led to similar results, we found that neglecting to include the ion gel effects in the transfer matrix model would lead to a slight underestimation of the scattering rate. To verify that the presence of the ion gel didn't change our assumptions in our extractions of the conductivity, we also repeated the calculation of $\sigma(\omega)$ via the Kramers-Kronig constrained variational analysis \cite{Kuzmenko2005} that included the numerical transfer matrix technique.

\section*{Data availability}

The datasets and materials used and/or analyzed during the current study are available from the corresponding authors on reasonable request.

\clearpage

\bibliography{main}{}

\section*{Acknowledgements}

We thank H. Bechtel and M. C. Martin for help with Raman characterization, H. Choi for contributions to early versions of the setup, and J. Heyman for helpful discussions. Ultrafast studies were supported by the U.S. Department of Energy, Office of Basic Energy Sciences, Materials Sciences and Engineering Division under contract DE-AC02-05CH11231 (Ultrafast Materials Science program KC2203). J.H.B. gratefully acknowledges a DFG postdoctoral fellowship. Work at NRL was support through Base programs via the Office of Naval Research. R.P.S. was partially supported by a Faculty Support Grant from the California State University - East Bay, Division of Academic Affairs.

\section*{Author contributions statement}

R.A.K. conceived the experiments; J.T.R., S.-F.S. and F.W. performed sample fabrication; G.C., R.P.S. and J.H.B conducted the experiments, G.C. and R.P.S. analysed the results.  All authors reviewed the manuscript.

\end{document}